\documentclass[aps, showpacs, showkeys,nofootinbib,floatfix]{revtex4}

\usepackage{amssymb}
\usepackage{amsmath}
\usepackage{graphicx}
\usepackage{hyperref}

\begin{document}
\hfill TP-DUT/2007-01V2

\title[Reconstruction of Deceleration Parameters from Recent Cosmic Observations]{Reconstruction of Deceleration Parameters from Recent Cosmic Observations}

\author{Lixin Xu\footnote{Corresponding author: lxxu@dl.cn}, Chengwu Zhang, Baorong Chang and Hongya Liu}

\address{School of Physics \& Optoelectronic Technology, Dalian
University of Technology, Dalian, 116024, P. R. China}


\begin{abstract}
In this paper, three kinds of simple parameterized deceleration
parameters $q(z)=a+\frac{b z}{1+z}$, $q(z)=a+\frac{b z}{(1+z)^2}$
and $q(z)=\frac{1}{2}+\frac{a z+b}{(1+z)^2}$ are reconstructed from
the latest Sne Ia Gold dataset, observational Hubble data and their
combination. It is found that the transition redshift from
decelerated expansion to accelerated expansion $z_T$ and current
decelerated parameter values $q_0$ are consistent with each other in
$1\sigma$ region by only using Sne Ia Gold dataset and observational
Hubble data in three parameterizations respectively. By combining
the Sne Ia Gold dataset and observational Hubble data together, a
tight constraint is obtained. With this combined constraints, $z_T$
is $0.505^{+0.080}_{-0.052}$, $0.368^{+0.059}_{-0.036}$,
$0.767^{+0.121}_{-0.126}$ with $1\sigma$ error in three
parameterizations respectively. And, it is easy to see that $z_T$
separates from each other in $1\sigma$ region clearly.
\end{abstract}

\maketitle

\section{Introduction}

At the end of the last decade, the observations of High redshift
Type Ia Supernova from two teams \cite{ref:Riess98,ref:Perlmuter99}
indicated that our universe is undergoing accelerated expansion.
Meanwhile, this suggestion was strongly confirmed by the
observations from WMAP
\cite{ref:Bernardis00,ref:Hanany00,ref:Spergel03,ref:Spergel06} and
Large Scale Structure survey \cite{ref:Tegmark}. To understand the
late-time accelerated expansion of the universe, a large part of
models are proposed by assuming the existence of an extra energy
component, dubbed dark energy, which has negative pressure and
dominates at late time to push the universe from decelerated
expansion to accelerated expansion. In principle, a natural
candidate to dark energy could be a small cosmological constant
$\Lambda$ which has the constant equation of state (EOS)
$w_{\Lambda}=-1$. However, there exist serious theoretical problems:
fine tuning and coincidence problems. To overcome the coincidence
problem, dynamical dark energy models, such as quintessence
\cite{ref:quintessence}, phantom \cite{ref:phantom}, k-essence
\cite{ref:k-essence}, Chaplygin gas \cite{ref:cha-gas}, holographic
dark energy \cite{ref:holo}, etc., are proposed.

Another approach to study the dark energy is by an almost
model-independent way, i.e., by a parameterized EOS of dark energy
which is implemented by giving a concrete form of the EOS of dark
energy directly, such as $w(z)=w_0+w_1 z$ \cite{ref:Cooray99},
$w(z)=w_0+w_1\frac{z}{1+z}$ \cite{ref:Cheallier00,ref:Lider03},
$w(z)=w_0+w_1\ln(1+z)$ \cite{ref:Gerke02}, etc.. Through this
method, the evolution of dark dark energy with respect to the
redshift $z$ is explored, and it is found that the current
constraints favor a dynamical dark energy, though the cosmological
constant is not ruled out in $1\sigma$ region \cite{ref:Riess06}.
Also, the dark energy favors a quintom-like dark energy, i.e. a
crossing of the cosmological constant boundary $w=-1$. In all, it is
an effective method to rule out the dark energy models. As known,
the universe is dominated by dark energy and is undergoing
accelerated expansion at present and was dominated by dark matter
and underwent a decelerated epoch in the past. In another words, the
universe underwent a transition from decelerated expansion to
accelerated expansion. So, to realize the transition, the
parameterized decelerated parameter is presented in a model
independent way by giving a concrete form of decelerated parameter
which is positive in the past and negative at present
\cite{ref:Banerjee05,ref:Xu06,ref:Gong06,ref:Alam2006}. Moreover, it
is interesting and important to know what is the transition time
$z_T$ from decelerated expansion to accelerated expansion. This is
the main point of this paper to be explored in the model-independent
way. In this paper, the Sne Ia Gold dataset and observational Hubble
data are used to constrain the transition redshift $z_T$ and current
value of decelerated parameter.

This paper is structured as follows. In section \ref{sec:II}, three
kinds of parameterized decelerated parameters are constrained by
latest $182$ Sne Ia Gold data points compiled by Riess
\cite{ref:Riess06} and observational Hubble data \cite{ref:SVJ2005}.
Section \ref{sec:III} is the conclusion.

\section{Reconstruction of Deceleration Parameter}\label{sec:II}

We consider a flat FRW cosmological model containing dark matter and
dark energy with the metric
\begin{equation}
ds^2=-dt^2+a^2(t)dx^2.
\end{equation}
The Friedmann equation of the flat universe is written as
\begin{equation}
H^2=\frac{8\pi G}{3}\left(\rho_{m}+\rho_{de}\right),
\end{equation}
where, $H\equiv \dot{a}/a$ is the Hubble parameter, and its
derivative with respect to $t$ is
\begin{equation}
\dot{H}=\frac{\ddot{a}}{a}-\left(\frac{\dot{a}}{a}\right)^2,
\end{equation}
which combined with the definition of the deceleration parameter
\begin{equation}
q(t)=-\frac{\ddot{a}}{aH^2},
\end{equation}
gives
\begin{equation}
\dot{H}=-\left(1+q\right)H^2.\label{dotH}
\end{equation}
By using the relation $a_0/a=1+z$, the relation of $H$ and $q$, {\it
i.e.}, Eq. (\ref{dotH}) can be written in its integration form
\begin{equation}
H(z)=H_0\exp\left[\int_{0}^{z}\left[1+q(u)\right]d\ln(1+u)\right],
\end{equation}
where the subscript "$0$" denotes the current values of the
variables. If the function of $q(z)$ is given, the evolution of the
Hubble parameter is obtained. In this paper, we consider three kinds
of parameterized deceleration parameters:
\begin{itemize}
\item A. $q(z)=a+\frac{b z}{1+z}$,\qquad $H(z)=H_0\left(1+z\right)^{1+a+b}\exp\left(-\frac{bz}{1+z}\right)$
\item B. $q(z)=a+\frac{b z}{(1+z)^2}$,\qquad $H(z)=H_0\left(1+z\right)^{1+a}\exp\left(\frac{bz}{2(1+z)^2}\right)$
\item C. $q(z)=\frac{1}{2}+\frac{a z+b}{(1+z)^2}$ \cite{ref:Gong06},\qquad $H(z)=H_0\left(1+z\right)^{3/2}\exp\left[\frac{b}{2}+\frac{a
z^2-b}{2(1+z)^2}\right]$
\end{itemize}
where, $a$, $b$ are constants which can be determined from the
cosmic observations, such as Sne Ia and observational Hubble data.
From the explicit expressions of Hubble parameters, this mechanisms
can also be treated as parameterizations of Hubble parameters which
can been constrained from observational Hubble data directly.

Before constraining the parameterizations of decelerated parameters,
we give brief discussion on the parameterized equations. It is easy
to obtain the current values of decelerated parameters which
determined at the redshift $z=0$ are $q_{0A}=a$, $q_{0B}=a$ and
$q_{0C}=1/2+b$ in three parameterizations A, B and C respectively.
Also, the transition redshift $z_T$ from decelerated expansion to
accelerated expansion can also be obtained by solving the equation
of $q(z=z_T)=0$. Then, they can be written as a function $z_T=z_T(a,
b)$ in terms of parameters $a$ and $b$ uniformly, if the equation
has real root.

Now, these models can be constrained by the Sne Ia Gold dataset and
observational Hubble data. The Sne Ia Gold dataset contains $182$
Sne Ia data \cite{ref:Riess06} by discarding all Sne Ia with
$z<0.0233$ and all Sne Ia with quality='Silver'. Constraints from
Sne Ia can be obtained by fitting the distance modulus $\mu(z)$
\begin{equation}
\mu_{th}(z)=5\log_{10}(D_{L}(z))+\mathcal{M},
\end{equation}
where, $D_{L}(z)$ is the Hubble free luminosity distance $H_0
d_L(z)$ and
\begin{eqnarray}
d_L(z)&=&(1+z)\int_{0}^{z}\frac{dz^{\prime}}{H(z^{\prime})}\\
\mathcal{M}&=&M+5\log_{10}\left(\frac{H_{0}^{-1}}{Mpc}\right)+25
\nonumber\\
&=&M-5\log_{10}h+42.38,
\end{eqnarray}
where, $M$ is the absolute magnitude of the object (Sne Ia here).
With Sne Ia dataset, the best fit values of parameters in dark
energy models can be determined by minimizing
\begin{equation}
\chi_{SneIa}^2(p_s)=\sum_{i=1}^{N}\frac{\left(\mu_{obs}(z_i)-\mu_{th}(p_s;z_i)\right)^2}{\sigma^2_{i}},
\end{equation}
where $N=182$ for Sne Ia Gold dataset, $\mu_{obs}(z_i)$ is the
moduli obtained from observations, $\sigma_{i}$ is the total
uncertainty of the Sne Ia data, and $p_s$ denotes the parameters
contained in the model. The Sne Ia datasets are used as cosmic
constraints can also be found in \cite{ref:constraints}.

The observational Hubble data are based on differential ages of the
galaxies \cite{ref:JL2002}. In \cite{ref:JVS2003}, Jimenez {\it et
al.} obtained an independent estimate for the Hubble parameter using
the method developed in \cite{ref:JL2002}, and used it to constrain
the EOS of dark energy. The Hubble parameter depending on the
differential ages as a function of redshift $z$ can be written in
the form of
\begin{equation}
H(z)=-\frac{1}{1+z}\frac{dz}{dt}.
\end{equation}
So, once $dz/dt$ is known, $H(z)$ is obtained directly
\cite{ref:SVJ2005}. By using the differential ages of
passively-evolving galaxies from the Gemini Deep Deep Survey (GDDS)
\cite{ref:GDDS} and archival data \cite{ref:archive}, Simon {\it et
al.} obtained $H(z)$ in the range of $0\lesssim z \lesssim 1.8$
\cite{ref:SVJ2005}. The observational Hubble data from
\cite{ref:SVJ2005} are list in Table \ref{Hubbledata}.
\begin{table}[htbp]
\begin{center}
\begin{tabular}{c|lllllllll}
\hline\hline
 $z$ &\ 0.09 & 0.17 & 0.27 & 0.40 & 0.88 & 1.30 & 1.43
 & 1.53 & 1.75\\ \hline
 $H(z)\ ({\rm km~s^{-1}\,Mpc^{-1})}$ &\ 69 & 83 & 70
 & 87 & 117 & 168 & 177 & 140 & 202\\ \hline
 $1 \sigma$ uncertainty &\ $\pm 12$ & $\pm 8.3$ & $\pm 14$
 & $\pm 17.4$ & $\pm 23.4$ & $\pm 13.4$ & $\pm 14.2$
 & $\pm 14$ &  $\pm 40.4$\\
\hline
\end{tabular}
\end{center}
\caption{\label{Hubbledata} The observational $H(z)$
data~\cite{ref:SVJ2005,ref:JVS2003} (see~\cite{ref:SR2006,ref:WZ}
also).}
\end{table}

The best fit values of the model parameters from observational
Hubble data \cite{ref:SVJ2005} are determined by minimizing
\begin{equation}
\chi_{Hub}^2(p_s)=\sum_{i=1}^9 \frac{[H_{th}(p_s;z_i)-H_{
obs}(z_i)]^2}{\sigma^2(z_i)}
\end{equation}
where $p_s$ denotes the parameters contained in the model, $H_{th}$
is the predicted value for the Hubble parameter, $H_{obs}$ is the
observed value, $\sigma(z_i)$ is the standard deviation measurement
uncertainty, and the summation is over the $9$ observational Hubble
data points at redshifts $z_i$. In our three cases, the derived
$H_{th}$ contains parameter $H_0$ which is current value of Hubble
parameter, and $H_0=72\pm2$ are taken as a prior.

\begin{figure}[tbh]
\centering
\includegraphics[width=5.6in]{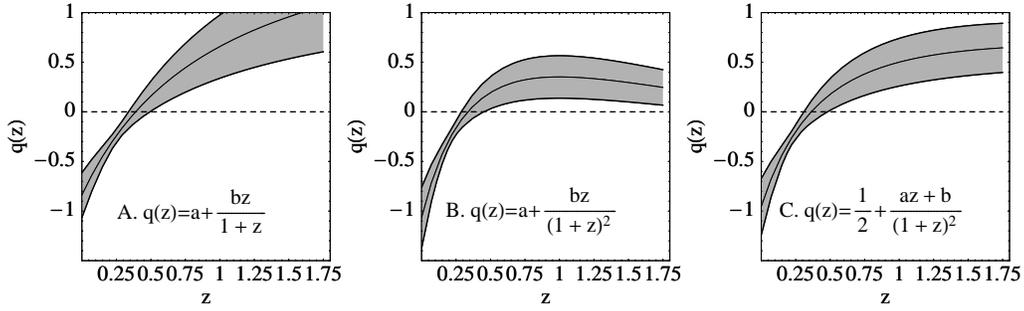}
\caption{The evolution of decelerated parameters with respect to the
redshift $z$, where the parameters $a$ and $b$ are determined by
fitting $182$ Sne Ia Gold dataset. The center solid lines are
plotted with the best fit values respectively in A, B and C, where
the shadows denote the $1\sigma$ regions.}\label{fig:snqz}
\end{figure}

Fitting the $182$ Sne Ia Gold data only, the minimum $\chi^2$ and
the best fit parameters $a$, $b$ and the transition times (or
redshift) $z_T$ in these three kinds of parameterizations are listed
in Table \ref{tab:snresults}. The evolutions of the decelerated
parameters $q(z)$ with $1\sigma$ errors are plotted in Fig.
\ref{fig:snqz}.

\begin{table}[tbh]
\begin{center}
\begin{tabular}{c|c|c|c|c|c}
\hline\hline
Parameters  & $\chi^2_{min}$ & $a$ & $b$ & $z_T$ & $q_0$\\
\hline A. $q(z)=a+\frac{b z}{1+z}$ & $156.44$ &
$-0.84^{+0.22}_{-0.22}$ & $3.00^{+1.05}_{-1.05}$ &
$0.39^{+0.10}_{-0.05}$ & $-0.84^{+0.22}_{-0.22}$\\ \hline B.
$q(z)=a+\frac{b z}{(1+z)^2}$ & $156.71$ & $-1.07^{+0.30}_{-0.30}$ &
$5.68^{+2.00}_{-2.00}$ & $0.34^{+0.11}_{-0.05}$ &
$-1.07^{+0.30}_{-0.30}$
\\ \hline
C. $q(z)=\frac{1}{2}+\frac{a z+b}{(1+z)^2}$ & $156.54$ & $1.46^{+1.22}_{-1.22}$ & $-1.46^{+0.28}_{-0.28}$ & $0.36^{+0.12}_{-0.05}$ & $-0.94^{+0.28}_{-0.28}$\\
\hline
\end{tabular}
\caption{The best fit results with $1\sigma$ error of constraints
from $182$ Sne Ia Gold dataset.}\label{tab:snresults}
\end{center}
\end{table}

\begin{figure}[tbh]
\centering
\includegraphics[width=5.6in]{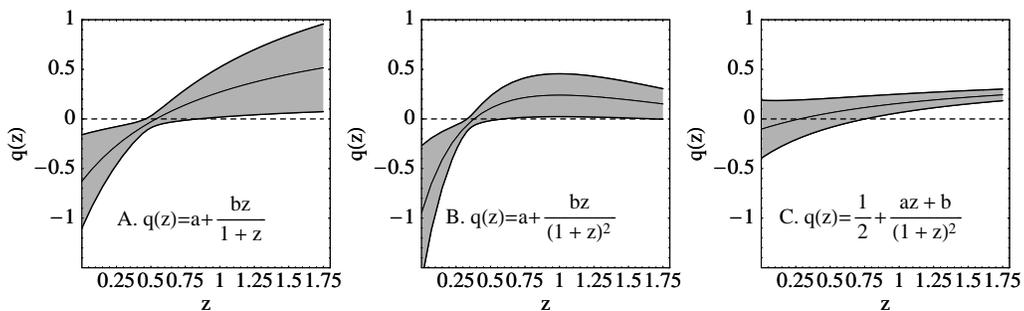}
\caption{The evolution of decelerated parameters with respect to the
redshift $z$, where the parameters $a$ and $b$ are determined from
the fitting of observational Hubble dataset. The center solid lines
are plotted with the best fit values respectively in A, B and C,
where the shadows denote the $1\sigma$ regions.}\label{fig:hubbleqz}
\end{figure}

By fitting the dataset from $9$ points of observational Hubble data
alone, we obtain the results of the minimum $\chi^2$ and the best
fit parameters $a$, $b$ and the transition times (redshift) $z_T$
which are listed in Table \ref{tab:hubbleresults}. The evolutions of
the decelerated parameters are plotted in Fig. \ref{fig:hubbleqz}.

\begin{table}[tbh]
\begin{center}
\begin{tabular}{c|c|c|c|c|c}
\hline\hline
Parameters  & $\chi^2_{min}$ & $a$ & $b$ & $z_T$ & $q_0$\\
\hline A. $q(z)=a+\frac{b z}{1+z}$ & $4.34$ &
$-0.63^{+0.47}_{-0.47}$ & $1.80^{+1.42}_{-1.42}$
&$0.54^{+0.30}_{-0.08}$ & $-0.63^{+0.47}_{-0.47}$\\ \hline B.
$q(z)=a+\frac{b
z}{(1+z)^2}$ & $4.07$ & $-0.95^{+0.68}_{-0.68}$ & $4.75^{+3.53}_{-3.53}$ & $0.38^{+0.20}_{-0.05}$ & $-0.95^{+0.68}_{-0.68}$\\
\hline
C. $q(z)=\frac{1}{2}+\frac{a z+b}{(1+z)^2}$ & $4.28$ & $-0.77^{+0.10}_{-0.10}$ & $-0.60^{+0.29}_{-0.29}$ & $0.28^{+0.47}_{-null}$ & $-0.10^{+0.29}_{-0.29}$\\
\hline
\end{tabular}
\caption{The best fit results from $9$ observational Hubble data.
Here '$null$' denotes the absence of transition from decelerated
expansion to accelerated expansion in $1\sigma$ low bound in case C
which can be seen in the right panel of Fig. \ref{fig:hubbleqz}
clearly.}\label{tab:hubbleresults}
\end{center}
\end{table}

In the above, three kinds of parameterized decelerated parameters
have been constrained by Sne Ia Gold dataset and observational
Hubble data respectively. However, the cosmological parameters have
degeneracies in almost all cosmological observables. So, it is
necessary to combine all available probes or observations to break
the degeneracies to obtain tight constraints. In the remained parts
of this section, the Sne Ia Gold dataset and observational Hubble
data are combined together to give a tight constraint to
parameterized decelerated parameters.

\begin{figure}[tbh]
\centering
\includegraphics[width=5.6in]{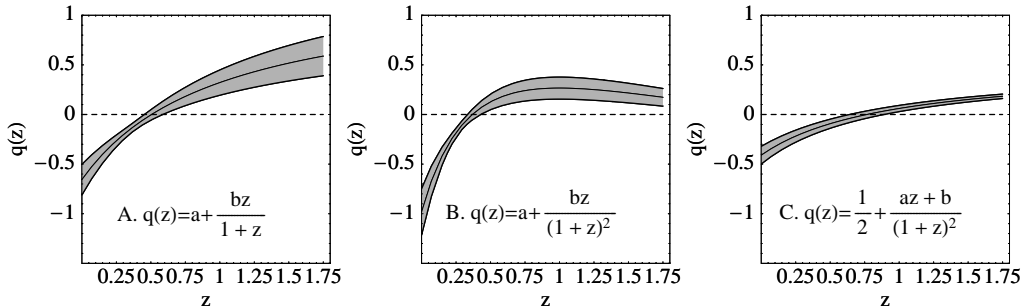}
\caption{The evolution of decelerated parameters with respect to the
redshift $z$, where the parameters $a$ and $b$ are determined from
the combined constraints from $182$ Sne Ia Gold dataset and $9$
observational Hubble data. The center solid lines are plotted with
the best fit values respectively in A, B and C, where the shadows
denote the $1\sigma$ regions.}\label{fig:ghqz}
\end{figure}

The best fit parameter values of $a$, $b$ and $z_T$ can be obtained
by minimizing the summation of $\chi^2$ of Sne Ia Gold dataset and
Hubble parameter data in these three kinds of parameterizations
\begin{equation}
\chi^2_{total}(p_s)=\chi_{SneIa}^2(p_s)+\chi_{Hub}^2(p_s).
\end{equation}
The results of combined constraints are listed in Table
\ref{tab:results}. The evolutions of the decelerated parameters
$q(z)$ with $1\sigma$ error are plotted in Fig. \ref{fig:ghqz}.
\begin{table}[tbh]
\begin{center}
\begin{tabular}{c|c|c|c|c|c}
\hline\hline
Parameters & $\chi^2_{min}$ & $a$ & $b$ & $z_T$ & $q_0$\\
\hline A. $q(z)=a+\frac{b z}{1+z}$ & $162.19$ &
$-0.657^{+0.153}_{-0.153}$ & $1.956^{+0.535}_{-0.535}$ &
$0.505^{+0.080}_{-0.052}$ & $-0.657^{+0.153}_{-0.153}$\\ \hline B.
$q(z)=a+\frac{b z}{(1+z)^2}$ & $161.01$ & $-0.982^{+0.232}_{-0.232}$
& $4.992^{+1.319}_{-1.319}$ & $0.368^{+0.059}_{-0.036}$ &
$-0.982^{+0.232}_{-0.232}$
\\ \hline
C. $q(z)=\frac{1}{2}+\frac{a z+b}{(1+z)^2}$ & $165.89$ & $-0.849^{+0.069}_{-0.069}$ & $-0.910^{+0.089}_{-0.089}$ & $0.767^{+0.121}_{-0.126}$
& $-0.410^{+0.089}_{-0.089}$\\
\hline
\end{tabular}
\caption{The best fit results of the combined constraints from $182$
Sne Ia Gold dataset and $9$ observational Hubble
data.}\label{tab:results}
\end{center}
\end{table}

\section{Discussion and Conclusion}\label{sec:III}

In this paper, by a model-independent way, we have used three kinds
of parameterized decelerated parameters to obtain the transition
time or redshift $z_T$ from decelerated expansion to accelerated
expansion and the current value of decelerated parameter $q_0$. To
obtain the best fit values of the transition redshift and
decelerated parameters, Sne Ia Gold dataset, observational Hubble
data and their combination are used as cosmic observational
constraints. With these cosmic observations, three parameterized
decelerated parameters are reconstructed. In Fig. \ref{fig:snqz}
where only Sne Ia Gold dataset is used, it can be seen that three
kinds of parameterizations all have transitions from decelerated
expansion to accelerated expansion. Also, it is found that the
transition redshift $z_T$ and current value of decelerated parameter
$q_0$ are consistent with each other in $1\sigma$ region. The
results may be caused by two possible reasons: a. the Sne Ia Gold
data points are not enough to discriminate them from each other; b.
three parameterizations just coincide with each other and have the
same transition redshift and current value of decelerated parameter.
In Fig. \ref{fig:hubbleqz} where observational Hubble data is used
alone, the similar results as shown in Fig. \ref{fig:snqz} can also
be obtained. However, the current values of decelerated parameter
have larger $1\sigma$ intervals which could come from the relative
lack of observational Hubble data points. Noticeably, in Fig.
\ref{fig:hubbleqz} (the right panel) the transition from decelerated
expansion to accelerated expansion is not clear in $1\sigma$ region
in parameterized C case where only observational Hubble data is
used. But, the best fit curve across the boundary of $q=0$. This
means that a different result will be obtained by choosing different
datasets as cosmic constraints. To give a tight constraint to the
transition redshifs and current values of decelerated parameter of
three kinds of parameterizations, a combined constraint of Gold Sne
Ia and observational data is introduced. The results are shown in
Table. \ref{tab:results} and Fig. \ref{fig:ghqz} where the combined
constraint is used, it can be seen that the transition redshifts
$z_T$ and current values of decelerated parameter $q_0$ separate
from each other clearly in $1\sigma$ region, i.e., three kinds of
parameterizations do not overlap with each other in $1\sigma$
regions. This means that the results rely on the concrete forms of
the parameterized equations strongly just as pointed out in Ref.
\cite{ref:BCK2004,ref:ST2006}, namely parameterization dependence.
The parameterization dependence is the potential drawback of
parameterized decelerated parameter and parameterized EOS of dark
energy. Now, a new question would be asked, which one parameterized
decelerated parameters describe our universe evolution or no one? To
answer this question completely, only the Sne Ia Gold dataset and
observational Hubble data are far from enough and another cosmic
observational constraints would be added. At last, it is worth to
remind that the chosen of cosmic observational datasets has a strong
impact on the results, as shown in Fig.
\ref{fig:snqz}-\ref{fig:ghqz}, not only in shrinking the error
regions but also in different best fit values, this is so-called
dataset dependence. In this paper, about the datasets of Sne Ia, the
Gold dataset is used only. It would be interesting to use SNLS
dataset as another constraint. And, we leave the comparison of Sne
Ia Gold and SNLS datasets as cosmic constraints in the future work.

\acknowledgements{This work is supported by NSF (10573003), NSF
(10647110), DUT (893321) and NBRP (2003CB716300) of P.R. China.}

\end{document}